\documentclass[aps, onecolumn,secnumarabic,amssymb,notitlepage,nobibnotes, nofootinbib,10pt]{revtex4-1}
\usepackage{graphicx}
\usepackage[pdftex,bookmarks,colorlinks,breaklinks]{hyperref}  
\usepackage{bm}
\usepackage{amsmath}
\begin{document}
\title{ A transient phase in cosmological evolution : A multi-fluid approximation for a quasi-thermodynamical equilibrium}
\author{Bob Osano$^{1,2}$}
\author{Timothy Oreta$^{1}$}
\affiliation{$^{1}$Cosmology and Gravity Group, Department of Mathematics and Applied Mathematics, University of Cape Town ($UCT$), Rondebosch 7701, Cape Town, South Africa \\} 
\affiliation{$^{2}$ Centre for Higher Education Development ($CHED$), University of Cape Town ($UCT$), Rondebosch 7701, Cape Town, South Africa }
\begin{abstract} This article presents the study of a multi-species fluid characterised by a freeze-out or break-away of one or more species. Whereas single-fluid approximation suffices for modelling of the pre {\it freeze-out} period, multi-fluid approximation is required for the post freeze-out period. Embedded in this is a transition period where neither one of the two approximations is singly appropriate.  We examine the thermodynamics of this fluid and find that the second law holds before, during and after the freeze-out.  Our application to cosmological modelling involving interacting dark-section species indicates that the species' equations of state are modified. \nopagebreak
\end{abstract}

\date{Received: date / Accepted: date}

\maketitle


\section{Introduction}
Cosmology is dominated by studies of segmented or {\it discretized} history of the universe or eras, whose dynamics are characterised by the dominant material. Examples of these include inflation \cite{Lin,Lid}, radiation\cite{Ber}, matter\cite{Ryd} and dark energy\cite{Frie}. Such studies are generally carried out using the single-fluid approximation in the sense that the modelling is based on a single observer world-line. To a large extent, cosmological observations have yielded results that agree with the predictions of the {\it standard Model} as seen, for example, in the analyses of the {\it cosmic microwave background} (CMB) radiation and the anisotropy thereof \cite{ade}. Although not all predictions are confirmed, what has been achieved has enabled us to build a probable-model of the evolving universe based on the scaffolding of the knowledge of the different eras. Nevertheless, the interplay between cosmological theory (theories) and observations have not always been smooth, resulting in several unanswered questions. Some observations lead to questions that demand the re-examination of the underlying theory or theories. Examples of these are "{\it axis of evil}" in the CMB \cite{Land} and late time acceleration\cite{Rie,Per}, just to mention two. The first line of attack has been to tweak the existing theory or improve technology with the hope that this could help explain the anomalous observation. Such attempts have had limited success, forcing some to suggest that there is a need for a complete overhaul of the underlying theories; whether of gravity or the material content of the universe. But there remains a yet to be explored alternative approach that may ease some of the tensions between theory and observation. The {\it modelling of transition} between eras does not feature much in literature but has the potential to resolve some of these issues. Such modelling applies to studies of epochs, where we think of a cosmological epoch as the event signifying when a change has taken place to an extent that it marks the beginning of a new era. As previously mentioned, the analysis is often performed assuming that the dynamics of the universe is dominated by one type of material, but a transition is required from one domination to the next. Even where the transition does not involve a switch in the dominant material, but a simple freeze-out, the dynamics of the flow is still impacted. 
This transition period is, predictably, complex and requires a completely different approach. The first line of attack is to assume that the transition is not instantaneous but occurs gradually allowing for a transient period that is not exclusively dominated by one of the fluid species. The purpose of this study is to model the transition between the different eras and to analyse the dynamics of such periods.
 
A study of multi-species fluids requires a way of approximating aggregated fluid properties, for example using the {\it multi-fluid approximation} on the one hand and on the other ways of dealing with how the different species interact. This may require ways of examining thermodynamics assuming the components interact thermodynamically. To carry out such a study, one needs to link theories built on single fluid dynamics and thermodynamics \cite{Prix,Eck,Lan,His} to those built on relativistic multi-fluids or multi-species fluids \cite{Isra0,Cart0,Priou} and their thermodynamics \cite{Isra1,Mull,Cart1,And1}. It may also require one to go beyond perfect fluids, and consider fluids that exhibit dissipation and those in which bulk viscosity plays a role \cite{Roy}.
This paper is a modest attempt at modelling the departure from the standard modelling of cosmological dynamics during the transition between epochs.

The paper is arranged as follows, section (\ref{sec1}) discusses single-fluid and multi-fluid approximation for a fluid with a break away component. Section (\ref{sec3}) discusses generalised second law of thermodynamics for a multi-fluid system, and section (\ref{sec4}) gives the general discussions and conclusions.

\section{\label{sec1}From Single-fluid to Multi- fluid approximation}
Consider fluid species denoted by $X, Y, Z$. Although we have three particle species, we have a single observer world-line and hence $t\equiv u^{a}\nabla_{a}$ where $u^{a}$ is the common 4-velocity. Let this 4-velocity be the determinant of a word-line of a fiducial {\it frame of reference}. This is a single-fluid approximation. The metric is given by
\begin{eqnarray}\label{MET}
ds^{2}=-dt^{2}+a^{2}(t)\left(\frac{dr^2}{1-\kappa r^2}+r^2 d\Omega^{2}\right),
\end{eqnarray} where $a(t)$ is the scale-factor such that $\dot{a}(t)/a(t)=3H$. $H$ is the Hubble parameter.
\begin{eqnarray}\label{Fried}
H^{2}+\frac{\kappa}{a^2}=\frac{1}{3}(\rho_{X}+\rho_{Y}+\rho_{Z}),
\end{eqnarray} where $\rho$ is the energy density. We have set $8\pi G=1=c^2$. It is easy to show that the corresponding Friedman equation is
\begin{eqnarray}\label{eq2}
\frac{d}{dt} \left(H^{2}+\frac{\kappa}{a^2} \right)=\frac{1}{3}(\dot{\rho}_{X}+\dot{\rho}_{Y}+\dot{\rho}_{Z}).
\end{eqnarray} This expresses the evolution of material content of the Universe in terms of hydrodynamic or single fluid approximation, and this formulated at a thermodynamical equilibrium state. We are interested in transitions between epochs and where the content of the Universe could be described as being in thermal quasi-equilibrium and where hydrodynamical approximation begins to break down. Whereas some transitions such as those before matter-radiation decoupling can be analysed using hydrodynamic approximation and thus using a single fluid in fluid in thermal equilibrium, there are others which require multi-species fluid approximation, while others will require a full multi-fluid approximation. For example, a transition that involves a species being 'frozen-out' when timescales become comparable to the timescale of the cosmic expansion could lead to the species breaking from the equilibrium. An example of this is the radiation dominated epoch and just before decoupling but after nucleosynthesis but where the material content is made up of plasma of nucleons, electrons and photons ( all in thermal equilibrium in which some of the content interact via radiative processes like Thompson scattering) and ends with baryons and electrons becoming separate fluids after breaking away from the thermal equilibrium. Only photon gas is left as a remnant of the earlier cosmic plasma. Whereas the initial nucleons, electrons and photons fluid can be treated as single fluid, baryon and electrons are best treated using multi-species fluid approximation\footnote{The authors would like to thank the anonymous reviewer for pointing this out}. The treatment of such {\it break-a-way} behaviour is what we would like to model. The conservation equations for the individual species energy-densities are given by
\begin{eqnarray}\label{densitydot}
\dot{\rho}_{X}&+&3H(\rho_{X}+p_{X})=-Q\nonumber\\
\dot{\rho}_{Y}&+&3H(\rho_{Y}+p_{Y})=Q\nonumber\\
\dot{\rho}_{Z}&+&3H(\rho_{Z}+p_{Z})=0,
\end{eqnarray} where $\mathcal{Q}$ \cite{Jamil} is the interaction term. The crucial aspect of the above conservations is the assumption that the species are in thermodynamics equilibrium therefore evolving together and single-fluid approximation therefore applies. What if one of the species breaks away? In particular let species $Y$ be made of two subspecies $Y_{1}$ and $Y_{2}$, where only $Y_{1}$ interacts with species $X$ and where subspecies $Y_{2}$ is able to break away from equilibrium. This may be presented as follows:
\begin{eqnarray}\label{densitydot1}
\dot{\rho}_{X}&+&3H(\rho_{X}+p_{X})=-Q\nonumber\\
\dot{\rho}_{Y_{1}}&+&3H(\rho_{Y_{1}}+p_{Y_{1}})=Q\nonumber\\
\dot{\rho}_{Z}&+&3H(\rho_{Z}+p_{Z})=0,
\end{eqnarray} which are in a new or adjusted equilibrium and
\begin{eqnarray}
\dot{\rho}_{Y_{2}}&+&3H(\rho_{Y_{2}}+p_{Y_{2}})=0
\end{eqnarray} which is out of equilibrium with the first three. We emphasise that the equilibrium experienced by species in Eq. (\ref{densitydot1}) is different to that experienced by the species in Eq. (\ref{densitydot}). The readjustment to new equilibria is a process that is preceded by a quasi-equilibrium period requiring irreversible theory for the analysis of the multi-fluid dynamics. To the best of our knowledge, no full fledged version of such theory exists and will need to be developed. However, the extended irreversible thermodynamics theory or the {\it Mueller-Israel-Stewart (MIS) theory} (see appendix (\ref{miss})) offers a starting point for such a development and will be pursued elsewhere \cite{Bob0}. An approximation of such a development will suffice for this study. 

\section{\label{sec3}Generalised second law of thermodynamics for a multi-species fluid system}
In this section, we focus on a system of fluids made up of three species; dark energy $DE$, matter $M$ ( baryonic matter and non-baryonic dark matter) and radiation $\chi$. In particular, the generic species placeholders given in the previous section will be replaced as follows $X\equiv DE, Y\equiv M, Z\equiv\chi$, in which case $i=DE, M,\chi$. We note that the interaction involves the dark sector components only. It is known that perfect fluids in equilibrium state do not generate entropy or heating due to friction as their dynamics is devoid of dissipation and is reversible. However, perfect fluid models are inadequate for modelling most astrophysical and cosmological processes. Such processes are best modelled using more physically motivated realistic fluids which exhibit irreversible properties. Indeed, some processes in astrophysics and cosmology can only be understood as dissipative processes thereby requiring a relativistic theory of dissipative fluids\cite{Roy}. It has been shown that for single-fluid approximation, irreversible thermodynamics implies that the entropy is no longer conserved but grows per the second law of thermodynamics. We need to examine if the law holds in our multi-fluid approximation.

We consider the limit, in the evolution of the fluid, where the 'freeze-out' of one or more of the species just begins. Whereas we might have a dynamical apparent horizon just before the freeze-out, a Rindler horizon may be defined if the breakaway species were to exhibit a uniform acceleration or deceleration in comparison to the remaining species. It is in this 'freeze-out' transient period that we would like to apply the {\it MIS} theory. The application of this theory, which was initially developed for short-range interactions, to holographic Rindler horizon \cite{Rin} is speculative. Nevertheless, our knowledge of this regime is scant, which opens it up to speculation. In our context, the dynamical apparent horizon before the 'freeze-out' evolves into a Rindler-like horizon after the 'freeze-out'. It, therefore, makes sense to first consider the dynamical apparent horizon. Using spherical symmetry, the metric (\ref{MET}) can be expressed as
\begin{eqnarray}
ds^{2}=\gamma_{ab}dx^{a}dx^{b}+\tilde{r}^{2}d\Omega^{2}_{3},
\end{eqnarray}where $\tilde{r}=a(t)r$, $x^{0}=t$, $x^{1}=r$ and the 2D metric $\gamma_{ab}=$diag$(-1,~a^{2}/(1-kr^{2})$.
The dynamical apparent horizon is determined by the relation $\gamma^{ab}\partial_{a}\tilde{r}\partial_{b}\tilde{r}=0$ implying the vector $\nabla \tilde{r}$ is null on the apparent horizon surface. The apparent horizon radius for the FLRW \cite{Akbar} is
\begin{eqnarray}
\tilde{r}_{A}&=&\frac{1}{\sqrt{H^2+\frac{\kappa}{a^2}}}.
\end{eqnarray} This horizon can also be construed as a causal horizon \cite{Hay1,Hay2,Bak}. 

It follows from Eq. (\ref{Fried}) as shown in section (\ref{appara1}) that this radius evolves in times as,
\begin{eqnarray}
\label{rdot}\dot{\tilde{r}}_{A}&=&-\frac{\tilde{r}^3_{A}}{6}(\dot{{\rho}_{DE}}+\dot{{\rho}_{M'}}+\dot{{\rho}_{\chi}}),\end{eqnarray} for species in the reset equilibrium just after the freeze-ou, where $M'$ represents the remnant matter. It is also easy to show from, Eq. (\ref{densitydot}), that
\begin{eqnarray}
\label{rsum}\dot{\tilde{r}}_{A}&=&\frac{H\tilde{r}^3_{A}}{2}\sum_{i}\left({\rho}_{i}+p_{i}\right),
\end{eqnarray}where $i=DE, M, \chi$. At almost thermal-equilibrium, it follows that
\begin{eqnarray}
dS_{i}&=&\frac{1}{T_{(i)}}(p_{i}dV+dE_{i}-\mu_{i} dN_{i})\nonumber\\
dS&=&\sum_{i}\frac{1}{T_{(i)}}(p_{i}dV+dE_{i}-\mu_{i} dN_{i})\simeq\frac{1}{T}\sum_{i}(p_{i}dV+dE_{i}-\mu_{i} dN_{i}),
\end{eqnarray} where we define quasi-equilibrium by demanding the species temperature difference be negligible. In this regard, $T_{(i)}=T$. Of course this assumption is not necessary and the analysis of the full system can still be performed. However, the full detailed analysis is involved and will not be pursued here as it would distract from the primary goal of this study. We would like to examine how the total entropy evolves in time as mediated by the different species contributions in our quasi-equilibrium description.
The individual entropy evolution takes the form
\begin{eqnarray}
\dot{S}_{i}&=&\frac{1}{T}(4\pi\tilde{r}^{2}_{A}p_{i}\dot{\tilde{r}}_{A}+\dot{E}_{i}-\mu_{i}\dot{N}_{i})\nonumber\\
\end{eqnarray} The total entropy is given by $S=\sum_{i}S_{i}$ and can be viewed as $$S=S\left(S_{DE}, S_{M}, S_{\chi}\right).$$ We take the multi-fluid temperature, $T$, to be equal to the horizon temperature, $T_{h}$, as mediated by the geometry of the universe. It is known that most theories of gravity have both surface and bulk terms. The surface terms are often ignored in most of these theories when determining the field equation, yet when these surface terms are evaluated on the horizon they yield horizon entropy. There exists a holographic connection between the surface and the bulk terms \cite{Pad3,Mukh,Pad1,Pad4,Pad5}, and indirectly between horizon thermodynamics and space-time dynamics. To this end, we need time evolution of the volume,$V$, the internal energy, $E$, and the number density, $N$.
The volume and its evolution is given by
$V=\frac{4\pi \tilde{r}^{3}_{A}}{3}$ and $\dot{V}=4\pi\tilde{r}^{2}_{A}\dot{\tilde{r}}_{A}$ respectively. These are necessary in order to connect the thermodynamical
quantities such as the energies $E$ and pressures $P$, with the cosmological quantities, the energy densities $\rho$ and the pressures $p$.
The internal energy for the three species are
\begin{eqnarray}
E_{i}&=&\frac{4\pi}{3}\tilde{r}^{3}_{A}\rho_{i},
\end{eqnarray}where again $i=DE, M, \chi.$ It is clear from Euler's relation that
\begin{eqnarray}
p_{i}=p(\rho_{i}, s_{i}, n_{i}),
\end{eqnarray} where $\rho_{i}=E_{i}/V$, $s_{i}=S_{i}/V$ and $n_{i}=N_{i}/V$. If one were to ignore the transfer of energy due to the internal degrees of freedom but one, one could assume a barotropic equation of state consistent with adiabatic pressure, but we are interested in a much broader categorisation of fluid species and will there not implement this restriction in our analysis. It is known that the temperature of a horizon is related to its radius \cite{Jacob,Akbar,Pad4,Pad5,Cai} when black hole thermodynamics is extended to cosmology i.e.
\begin{eqnarray}
T_{h}&=&\frac{1}{2\pi\tilde{r}_{A}}
\end{eqnarray} The entropy of the horizon can be defined as $S_{h}=4\pi{\tilde{r}^{2}_{A}}/{4G}=8\pi^{2}\tilde{r}^{2}_{A}$, where $8\pi G=1$. The total entropy is then given by
\begin{eqnarray}
S_{Tot}\simeq\sum_{i}S_{i}+S_{h}.
\end{eqnarray}
The time evolution of the total entropy takes the form
\begin{eqnarray}
\dot{S}_{Tot}&\simeq&\dot{S}_{DE}+\dot{S}_{M'}+\dot{S}_{\chi}+\dot{S}_{h},
\end{eqnarray} where the overdot is the derivative with respect to cosmic time and where
\begin{eqnarray}
\dot{S}_{DE}&=&\frac{4\pi\tilde{r}^{2}_{A}}{T}\left[(\rho_{DE}+\rho_{DE})(\dot{\tilde{r}}_{A}-\tilde{r}_{A}H)-\frac{\tilde{r}_{A}}{3}\mathcal{Q}\right]-\frac{1}{T}\mu_{DE}\dot{N}_{DE} \nonumber\\
\dot{S}_{M'}&=&\frac{4\pi\tilde{r}^{2}_{A}}{T}\left[(\rho_{M'}+\rho_{M'})(\dot{\tilde{r}}_{A}-\tilde{r}_{A}H)+\frac{\tilde{r}_{A}}{3}\mathcal{Q}\right]-\frac{1}{T}\mu_{M'}\dot{N}_{M'} \nonumber\\
\dot{S}_{\chi}&=&\frac{4\pi\tilde{r}^{2}_{A}}{T}\left[(\rho_{\chi}+\rho_{\chi})(\dot{\tilde{r}}_{A}-\tilde{r}_{A}H)\right]-\frac{1}{T}\mu_{\chi}\dot{N}_{\chi}.
\end{eqnarray}
The horizon entropy evolves as $\dot{S}_{h}=16\pi^2\tilde{r}_{A}\dot{\tilde{r}}_{A}
$ and the total entropy therefore obeys the evolution equation
\begin{eqnarray}
\label{sTot}\dot{S}_{Tot}&\simeq &8\pi^2\tilde{r}^3_{A}\sum_{i}\left[(\rho_{i}+p_{i}))(\dot{\tilde{r}}_{A}-\tilde{r}_{A}H)\right]
-\sum_{i}2\pi\tilde{r}_{A}\mu_{i}\dot{N}_{i}+16\pi^{2}\tilde{r}_{A}\dot{\tilde{r}}_{A}.
\end{eqnarray} Using Eq.(\ref{rsum}),
\begin{eqnarray}
\label{sTot1}\dot{S}_{Tot}&\simeq &4\pi^2\tilde{r}^6_{A}H\left[\sum_{i}(\rho_{i}+p_{i})\right]^{2}-2\pi\tilde{r}_{A}\sum_{i}\mu_{i}\dot{N}_{i}
\end{eqnarray} It is clear that this finding holds regardless of the nature of gravitational interaction $\mathcal{Q}$. This result modifies the finding in \cite{Jamil} where the new equation of state for one of the species ( e.g. $DE$) emerges for the critical $\dot{S}_{Tot}=0,$ implying
\begin{eqnarray}
\left[\sum_{i}(\rho_{i}+p_{i})\right]^{2} &=&{\frac{1}{2\pi \tilde{r}^5_{A}H}\sum_{i}\mu_{i}\dot{N}_{i}}
\end{eqnarray} and on expanding the left hand-side
\begin{eqnarray}
\label{eos}\frac{p_{DE(cr)}}{\rho_{DE}}&=&-1-(\rho_{M'}+p_{M'})\frac{1}{\rho_{DE}}-(\rho_{\chi}+p_{\chi})\frac{1}{\rho_{DE}}\nonumber\\&&~~~+\frac{1}{\rho_{DE}}\sqrt{\frac{1}{2\pi \tilde{r}^5_{A}H}\sum_{i}\mu_{i}\dot{N}_{i}}.
\end{eqnarray} A similar analysis can be done for the extreme case were $DE$ or $DM$ or $\chi$ were to freeze-out. 

\section{\label{sec4}Discussions and Conclusions} Let's examine Eq. (\ref{sTot}). We know that $\tilde{r}_{A}, H,p_{i}, \rho_{i}$ are by definition positive. Eq. (\ref{rdot}) guarantees that $\dot{\tilde{r}}_{A}>0.$ The result of the summation will be positive since the horizon radius is greater than the Hubble parameter. This is confirmed by setting \begin{eqnarray}
(\rho_{i}+p_{i})(\dot{\tilde{r}}_{A}-\tilde{r}_{A}H)>0\end{eqnarray} which implies
\begin{eqnarray}
\frac{\dot{\tilde{r}}_{A}}{\tilde{r}_{A}}>H=\frac{\dot{a}}{a},
\end{eqnarray}as expected for the case where the surface term is neglected. Finally, $\dot{S}_{Tot}>0$ is achieved if $\sum_{i}\mu_{i}\dot{N}_{i}<0$, which is exactly what is expected of Gibbs free energy for negative chemical potentials. If we label Gibbs free energy using the letter $E_{G}$, then $\dot{E}_{G}<0$ implies $\dot{S}_{G}>0$. This establishes the generalised second law of thermodynamics for interacting dark-sector and at the onset freeze-out. We note that the inclusion of the chemical interaction in a multi-fluid approximation conserves the second law of thermodynamics. Effects of non-zero chemical potential on the equation of state of the dark energy in single-fluid approximation were examined in \cite{Lim} where it was found that the equation of state depends heavily on the magnitude and the sign of the chemical potential. Eq. (\ref{eos}) modifies that finding and has the potential to lift the $\omega_{cr}$ into the non-phantom state. We hesitate to provide an estimate as this would require the accurate estimation of $\tilde{r}_{A}$ and $\sum_{i}\mu_{i}\dot{N}_{i}$ in the quasi-equilibrium state. Using a multi-fluid approximation, we have investigated a cosmological scenario involving three particles species with two of these interacting both gravitationally and chemically.

We have examined the potential use of multifluid approximation to model fluid flow where one or more of the fluid species suffer 'freeze-out' thereby separating from the rest of the flow. For illustrative purposes, we considered the cosmological case involving dark-sector and radiation. The application is admittedly speculative, however, our scanty knowledge of the dark-sector opens room for such speculation. In this regard, we considered the universe as a thermodynamical system enclosed by the dynamical apparent horizon and calculated the separate entropy variation for each fluid species. The sum of these entropy variations together with that of the common horizon gives the total entropy of the universe. We find that the generalised second law of thermodynamics holds. It is important to note that we used dynamical apparent horizons and did not examine cases involving other types of horizons. Although we have illustrated that the generalised second law of thermodynamics in the interaction scenario involving dark energy ($DE$) and dark matter($M$) in the presence of radiation($\chi$), further investigation is still needed to make the findings applicable to quantitative or numerical cosmological analysis.
\appendix
\section{\label{miss} { \it The Mueller-Israel-Stewart (MIS)Theory}}
The {\it Mueller-Israel-Stewart (MIS)} theory was developed for single-fluid approximation \cite{Mull,Isra0,Isra1}.  the main parameters are the energy-momentum tensor, $T^{\mu\nu}$, and the particel 4-current, $N^{\nu}$, subject to the conservation laws
$\nabla_{\mu}T^{\mu\nu}=0,
\nabla_{\mu}N^{\mu}=0.$
 Given the 4-velocity, $u^{\mu}$, and the projection tensor, $h^{\mu\nu}=g^{\mu\nu}+u^{\mu}u^{\nu}$, the two state parameters decompose into
\begin{eqnarray}
N^{\mu}&=&{\bf n} u^{\mu}+n^{\mu}\\
\label{MIS}T^{\mu\nu}&=&\rho u^{\mu}u^{\nu}+Ph^{\mu\nu}+2u^{(\mu}q^{\nu)}+\pi^{\mu\nu},
\end{eqnarray}
where ${\bf n}=N^{\mu}u_{\mu}$ is the particle number density and $n^{\mu}$ is the diffusion current. $\rho=u_{\nu}u_{\mu}T^{\nu\mu}$ is the energy density, $P$ is the aggregate of the equilibrium, $ p$, and the bulk viscosity pressures, $\Pi$. The aggregate gives the isotropic pressure. The anisotropic pressure is given by $\pi^{\nu\mu}$ while $q^{\mu}$ is the heat flux vector. 
The extra quantities, $ q^{\mu},\pi^{\mu}, \Pi$ are the dissipative quantities. But this theory is for short range interactions within hydrodynamics regime. It's extension to multifluid, which extends beyond the hydrodynamics regime, is considered in \cite{Bob0}. We will refer to this as the $extended~MIS$ theory. In this extended theory, the primary extensive parameters are taken to be the species number flux current $N^{\mu}_{i}$, the stress momentum tensor $T^{\mu\nu}_{(i)}$, and the entropy flux vector $S^{\mu}_{(i)}$.  In this case, the total stress-energy momentum is conserved but not the individual i.e. $\nabla_{\mu}T^{\mu\nu}=\nabla_{\mu}(\sum_{i}T^{\mu\nu}_{(i)}+\bar{T}^{\mu\nu})=0\ne\nabla_{\mu}T^{\mu\nu}_{(i)}$, where $\bar{T}^{\mu\nu}$ is the momentum due to the interactions. The decompositions take the form;
\begin{eqnarray}
N^{\mu}_{(i)}={\bf n}_{(i)}u^{\mu}+n^{\mu}_{(i)},
\end{eqnarray}
where ${\bf n}_{i}$ is the species number density and $n^{\mu}_{i}$ is the species diffusion current. Similarly
\begin{eqnarray}
T^{\mu\nu}_{(i)}=\rho_{(i)}u^{\mu}u^{\nu}+p_{(i)}h^{\mu\nu}+2u^{(\mu}_{(i)}q^{\nu)}_{(i)}+\pi^{\mu\nu}_{(i)},
\end{eqnarray}
where $h^{\mu\nu}=g^{\mu\nu}+u^{\mu}u^{\nu}$ is the projection tensor to the rest frame of the various fluid species. This projection is only applicable upon the separation of species. Here too, $\rho_{(i)}=u_{\nu}u_{\mu}T^{\nu\mu}_{(i)}$ is the energy density, $p_{(i)}$ is the isotropic pressure, $\pi^{\nu\mu}_{(i)}$ is the anisotropic pressure, $q^{\mu}_{(i)}$ is the heat flux vector. The use of different velocities have also been considered in single-fluid approximation, for example in the comparison of the energy and the particle frames in \cite{Isra0}, or the rest frame and the boosted frame in \cite{Pad}. The entropy vector takes the form:

\begin{eqnarray}
S^{\mu}_{(i)}&=&s_{(i)}u^{\mu}+s^{\mu}_{(i)}\\&=&s_{(i)}u^{\mu}_{(i)}+\frac{q^{\mu}}{T}\nonumber\\&&-\left(\beta_{(i)}\Pi^2+\beta_{1(i)}q_{(i)\nu}q^{\nu}_{(i)}+\beta_{2(i)}\pi_{(i)\gamma\delta}\pi^{\gamma\delta}_{(i)}\right)\frac{u^{\mu}}{2T}\nonumber\\&&~~~~+\left(\alpha_{0(i)}\Pi q^{\mu}_{(i)}+\alpha_{1(i)}\pi^{\mu\nu}_{(i)}q_{(i)\nu}\right)\frac{1}{T}.\label{en1}
\end{eqnarray}
$s_{(i)}$ is the entropy density, $s^{\mu}_{(i)}$ is the entropy flux with respect to $u^{\mu}$ such that $s_{\mu(i)}u^{\mu}=0$. $\Pi_{(i)}$ is the bulk viscosity.
The complexity of the detailed interactions is immense but tractable as will be demonstrated in \cite{Bob0}. In this regard, the total entropy vector takes the phenomenological expressions
\begin{eqnarray}
S^{\mu}=\sum_{i}S^{\mu}_{i}+\bar{S}^{\mu},
\end{eqnarray} 
where, as in the stress-energy-momentum tensor, the term with the bar denotes interaction effects. It is important to reflect on the dynamics and changes that take place in this approximation. We postulate that there is a gradual change that sees the terms with the bars being important at $t_{qe}$ to having no effect at some $t>t_{qe}$. In other words, transiting to truly separate fluids requiring a full multifluid approximation. We now consider an application to cosmology. 
\section{An evolving apparent horizon}
We gave a definition of the horizon radius
\begin{eqnarray}\label{appara1}
\tilde{r}_{A}&=&\frac{1}{\sqrt{H^2+\frac{\kappa}{a^2}}}
\end{eqnarray} in section (\ref{sec1}). It will be noticed that one can replace the radius using Eq.(\ref{eq2}). In particular,
\begin{eqnarray}
H^2+\frac{\kappa}{a^2}&=&\frac{1}{\tilde{r}^{2}_{A}}=\frac{1}{3}(\rho_{DE}+\rho_{M}+\rho_{\chi}).
\end{eqnarray} The evolution with respect to proper time, $t$, yields
\begin{eqnarray}
\frac{d}{dt} \left(H^{2}+\frac{\kappa}{a^2} \right)&=&-2\frac{\dot{\tilde{r}}_{A}}{\tilde{r}^3_{A}}\nonumber\\&=&\frac{1}{3}(\dot{\rho}_{DE}+\dot{\rho}_{DM}+\dot{\rho}_{X}),\end{eqnarray} from which it follows that
\begin{eqnarray}
\dot{\tilde{r}}_{A}&=&-\frac{\tilde{r}^3_{A}}{6}(\dot{\rho}_{DE}+\dot{\rho}_{DM}+\dot{\rho}_{\chi})
\end{eqnarray}

\end{document}